\documentclass[preprintnumbers,amsmath,amssymb]{revtex4}
\usepackage{cases}
\usepackage{amsmath}
\usepackage{amssymb}
\usepackage{amsfonts}
\usepackage{amssymb}
\usepackage{dcolumn}
\usepackage{bm}
\usepackage{epsfig}
\usepackage{bbm}
\usepackage{graphicx}
\usepackage{xcolor}
\usepackage{array}
\usepackage{subfigure}
\usepackage{hyperref}
\usepackage{mathrsfs}
\usepackage{verbatim}
\usepackage{float}
\usepackage{epstopdf}
\usepackage{color}

\newcommand{\be}{\begin{equation}}
\newcommand{\ee}{\end{equation}}
\newcommand{\ba}{\begin{eqnarray}}
\newcommand{\ea}{\end{eqnarray}}

\newcommand{\gsim}{\mathrel{\hbox{\rlap{\lower.55ex \hbox {$\sim$}}
			\kern-.3em \raise.4ex \hbox{$>$}}}}
\newcommand{\lsim}{\mathrel{\hbox{\rlap{\lower.55ex \hbox {$\sim$}}
			\kern-.3em \raise.4ex \hbox{$<$}}}}

\hypersetup{colorlinks=true,
	breaklinks=true,
	pdfstartview=Fit,
	linkcolor=blue,
	citecolor=blue,
	urlcolor=blue}

\begin{document}

\title{Can sub-GeV dark matter coherently scatter on the electrons in the atom?}

\author{Ji-Heng Guo, Yu-Xuan Sun, Wenyu Wang, Ke-Yun Wu}
\email{
2944835554@emails.bjut.edu.cn, YuXuanSun@emails.bjut.edu.cn, wywang@bjut.edu.cn,\\ keyunwu@emails.bjut.edu.cn}
\affiliation{Faculty of Science, Beijing University of Technology, Beijing, China}

\begin{abstract}
A novel detection of sub-GeV dark matter is proposed in the paper.
The electron cloud is boosted by the dark matter 
and emits an electron when it is dragged back by the heavy nucleus,
namely the coherent scattering of the electron cloud of the atom.
The survey in the X-ray diffraction shows that the atomic form factors
are much more complicate than the naive consideration. The results
of the relativistic Hartree-Fock(RHF) method give non-trivial shapes of the atoms. The detailed calculation of the recoil of the electron cloud, the kinetics, the fiducial cross section and the corresponding calculation of detection rate are given analytically.
The numerical results show that the limits of the RHF form factors are 
much stringent than the recoil of a single electron,
almost 4 orders stronger, and also gives tight limitations comparing to the Migdal effect below about several hundred MeV. The physical picture and the corresponding results are promising and need 
further explorations.
\end{abstract}

%\pacs{95.30.Cq, 03.65.Nk}

\maketitle
\section{Introduction}
Observations in cosmology and astrophysics clearly support the presence of dark matter (DM)~\cite{Bertone:2004pz}. Its features, such as mass and interactions are however still unknown. One of the most promising experimental avenues is to search for small energy depositions from the dark matter elastically scattering in sensitive detectors on Earth. An attractive dark matter candidate is Weakly Interacting Massive Particles (WIMPs), which have been explored in various direct detection, indirect detection and collider experiments. The null results have produced very stringent limits on the WIMP-nucleus scattering cross section heavier than 1 GeV. Thus the hunt for sub-GeV dark matter is a hot topic at the cutting edge of physics research.Nevertheless, the sub-GeV dark matter can emerge in some fundamental theory 
such as the Next-to-Minimal supersymmetry standard model (NMSSM)~\cite{Abdughani:2017dqs,Cao:2015efs,Wang:2014kja}.
The self-interaction can be realized between the light dark matter to solve the small structure problems {\em etc}~\cite{Wang:2022lxn}. However, as the sensitivity is lost for the light dark matter detection, the traditional strategies for detecting WIMP-type dark matter are no longer feasible. Physicists are striving to create techniques for detecting light dark matter. 

Among all the new ways searching for the sub-GeV dark matter, the ionization signal on the detectors could be the most promising one.
In the dual-phase Xenon Time Projection Chamber, the Xenon atoms in the Liquid Xenon
phase can be ionized due to a collision. Then the ionized electrons drift into the
Gaseous Xenon layer at the top of the detector in presence of an external electric
field, which produces a proportional scintillation light, 
namely the S2 signal. In the theoretical studies, such ionization signals can
come from the DM-electron scattering~\cite{Essig:2011nj,
Essig:2012yx,Essig:2015cda,Chen:2015pha,Essig:2017kqs,
Cao:2020bwd,Bloch:2020uzh,Knapen:2021run,Hochberg:2021pkt,Xia:2021vbz,Xia:2020apm,Li:2014vza} or the DM-nucleus
scattering through the Migdal effect that originates from non-instantaneous 
movement of electron cloud during a nuclear recoil
event~\cite{Migdal:1939,Vergados:2005dpd,Moustakidis:2005gx,Ejiri:2005aj,
Bernabei:2007jz,Ibe:2017yqa,Dolan:2017xbu,Bell:2019egg,Baxter:2019pnz,Essig:2019xkx,
GrillidiCortona:2020owp,Liu:2020pat,Knapen:2020aky,Liang:2020ryg,Flambaum:2020xxo,
Bell:2021zkr,Acevedo:2021kly,keyun:2021}. 

In addition to those DM-electron scattering and the Migdal effect on the ionization
detection, there is another way which seems to be promising for the search of the
light dark matter. As it is well known that the electrons 
are bounded by the nucleus in an atom form the electron cloud.
If the dark matter can collide with the single electron in the atom, it can also
collide the whole cloud of the atom too. In fact, this is just the coherent
scattering in the dark matter detection. For example, there can be 
$Z^2$ (charge number) or $A^2$  (atom number) enhancements
in the spin-independent detection of the DM-nucleus scattering. As for 
the detection of the ionization, the question is that does
similar coherent enhancement exist in the recoils of the electron. 
This is not a trivial question since it can be seen in the X-ray
diffraction (XRD) on the atom, there do have the coherent scattering  
of electron cloud from which the atom scattering form factor can be derived.
As for the sub-GeV dark matter, the de Broglie wavelength of the dark matter
can be sufficiently small to inject into the target atom, the
same as the diffraction of the incident X-ray too.  The collision can 
boost the electron cloud to a sizable velocity. After that 
it can also be dragged back by the static heavy nucleus and 
throw away an electron to give similar ionization 
S2 signal. Such a mechanism seems reasonable, or at least should be checked that whether this effect can exist or can be significant for the detection.

In this paper, we will check this novel detection
of scattering on the electron cloud to see whether a coherent
enhancement exists in the recoils of the electron. As the process is similar to
the diffraction of X-ray scattering on the atom, we will show our understanding in
the coherent scattering from the point of the X-ray diffraction at first. Then
we will show how to calculate the coherent scattering between dark matter
and the electron cloud. After that, the constraints from the current
dark matter detection are calculated numerically for the comparison with
the results of the DM-electron scattering and the Migdal effect.
The paper is organized in the following: 
the coherent scattering of the electron in an atom are discussed 
in the sec.~\ref{sec2:cs}; the calculation methodology 
for the coherent scattering is given in the sec.~\ref{sec3:th};
the numerical results and the comparison with the other strategies are
shown in the sec.~\ref{sec4:nr} and our conclusion is given in the
sec.~\ref{sec5:con}.

\section{Coherent scattering of the electron in an atom}\label{sec2:cs}
\begin{figure}[htbp]  
{\includegraphics[width=5cm]{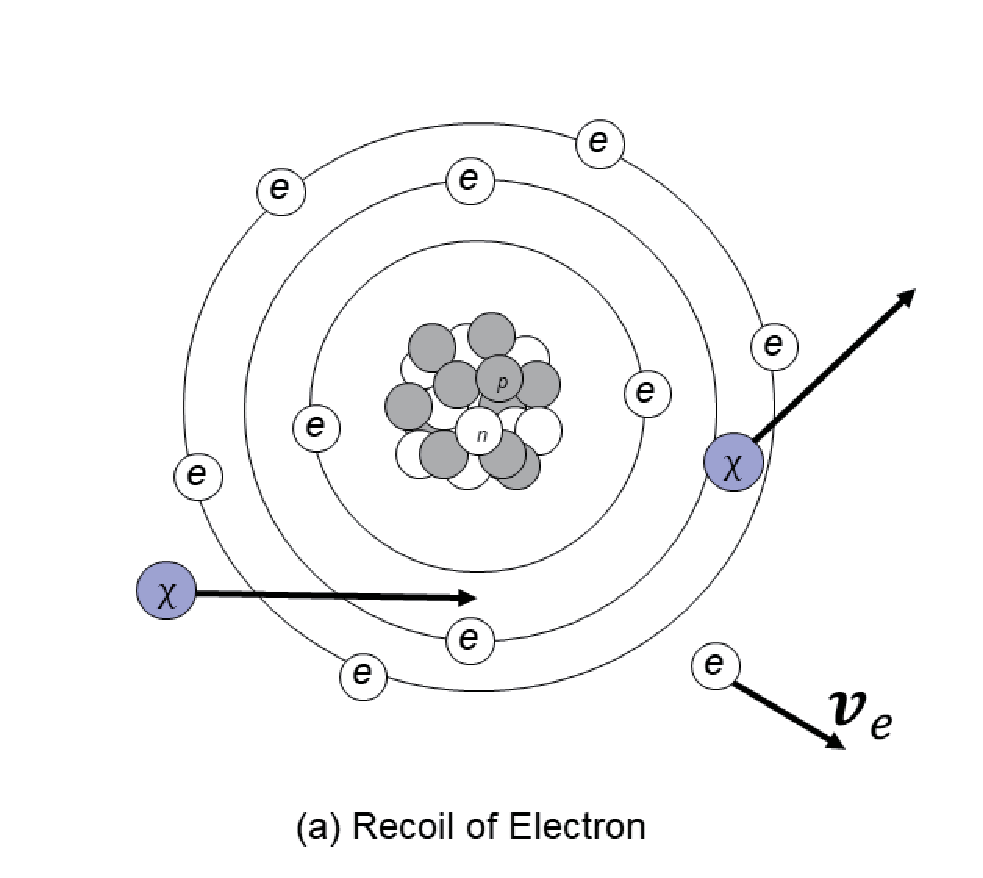}}~~~  
{\includegraphics[width=5cm]{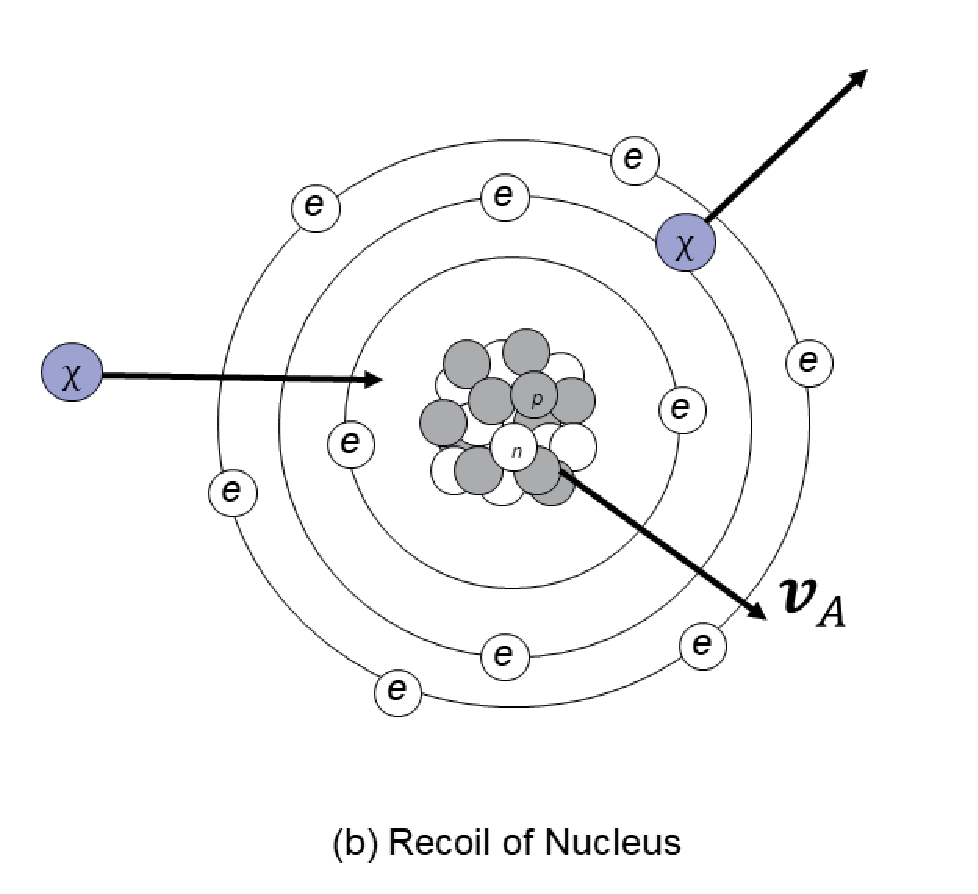}}~~~  
{\includegraphics[width=5cm]{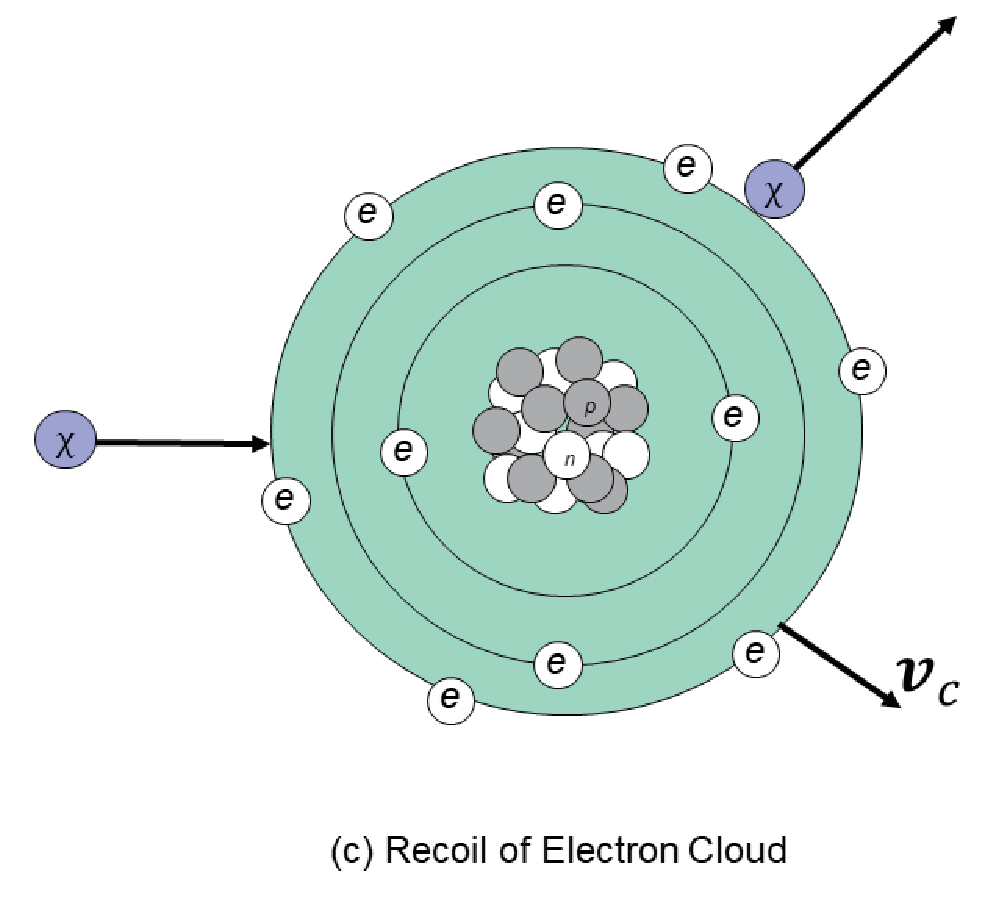}}  
\caption{The three processes which can induce ionization which are
the collision between dark matter and single electron, the Migdal effect and 
the collision between dark matter and the electron cloud of an atom, respectively.}
\label{fig1:cs}
\end{figure}
Fig.~\ref{fig1:cs} shows the scattering processes of the electrons when the
dark matter collide on the atom in which Fig.~\ref{fig1:cs} (a) shows  
sketch map of that the dark matter scattering on the single electron 
and Fig.~\ref{fig1:cs} (b) shows the sketch map of the Migdal effect.
Detailed calculation of the first two processes 
and the comparison  between them can be found in
Ref.~\cite{Baxter:2019pnz}. Here we just list essential points 
of these two processes:
\begin{enumerate}
    \item In the recoil of the electron, 
    the dark matter transfers momentum $q$ directly to the single electron
    bounded in the atom, giving the ionization electron with recoil energy $E_e$
    and momentum $k_e$. The impulse approximation\cite{Farina:1978} can be adopted in the theoretical calculation of the rate.
    \item For the Migdal effect, the dark matter transfer momentum $q$  to the nucleus. The momentum transfer happens much faster than the light-crossing time of the electron cloud, thus the nucleus get a sudden "movement" with velocity $v_A$ which can drag the electron cloud to same velocity and throw away an electron with energy $E_e$ and momentum $k_e$. The inner
    product of the containing phase between the eigenstate wave 
    function at rest and the Galilean transformation of the energy eigenstates of the moving atom gives the amplitude of the throwing rate of electron.
    \item The numerical results show that the Migdal effect is always
    subdominant to the electron scattering when the mediator is light. But it
    can dominate the ionization for heavy mediator and mass of the dark matter
    in hundreds of MeV range.
\end{enumerate}
In addition to these two processes, we propose another novel collision between
the dark matter and the electron cloud which is formed by the bounded 
electrons in an atom. The corresponding sketch map of the process is
shown  in the Fig.~\ref{fig1:cs} (c). The electron cloud is taken as
a ``ball" composed of the distributed electric charge, 
of which the amount is the charge number of the corresponding target element.
As the mass of the dark matter is at the MeV order, one can simply 
figure out the de Broglie wavelength of the dark matter is much smaller than
the radius of the target atom which is at the order of Bohr radius.
Thus the collision between dark matter and the electron cloud can be
considered as a collision between a ``point" and a ``puffy ball"
\cite{Chu:2020, Wenyu:2021}.
The corresponding scattering rate is always conveniently calculated
in the momentum space, and the Fourier transformation of the charge
distribution $\rho(r)$ gives the form factor $F(q)$ to account
for the coherence of the charge distribution. The whole picture
is similar to the diffraction of X-ray scattering on the crystal,
thus before going to further exploration the detection of the dark matter,
we discuss some essentials of X-ray diffraction.

\subsection{Atomic form factor in XRD}
\begin{figure}[htbp]
\begin{center}
{\includegraphics[width=12cm]{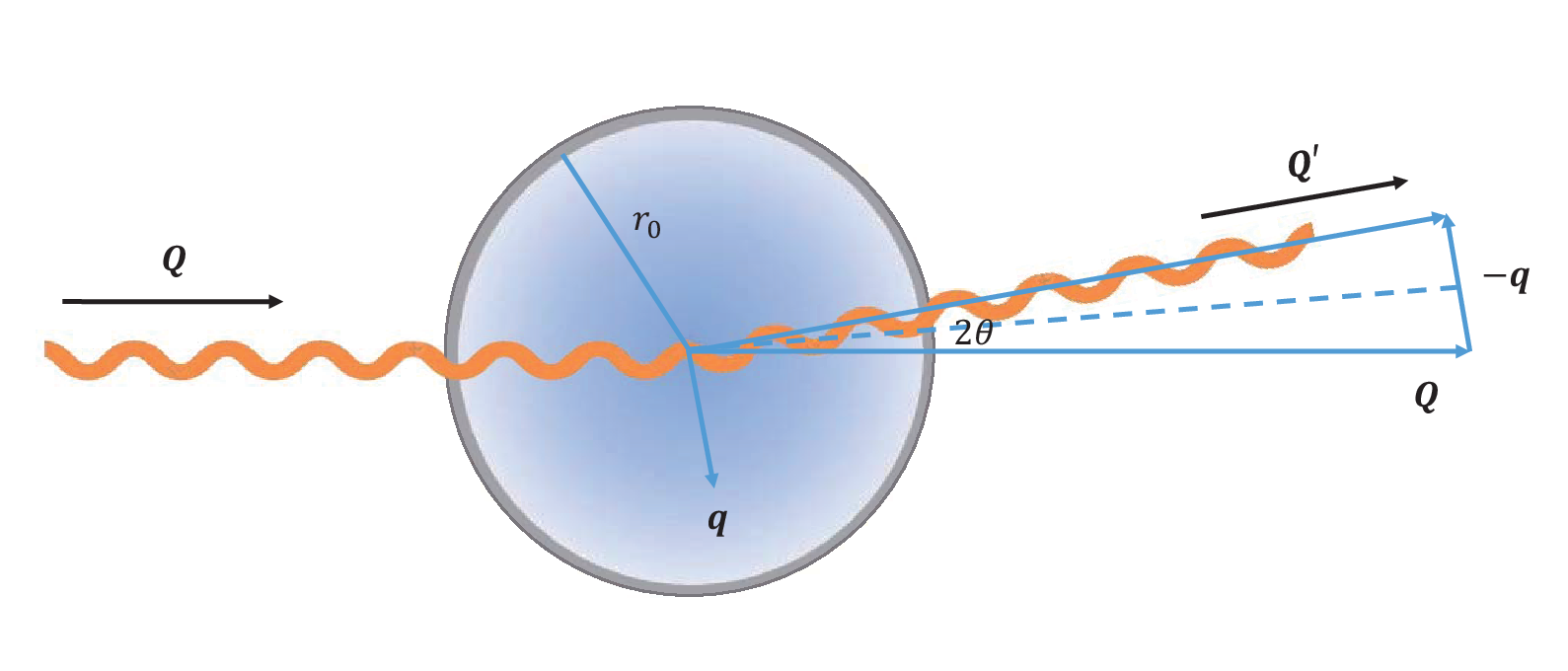}} 
\caption{The X-ray diffraction on the atom.} \label{fig2:xrd}
\end{center}
\end{figure}
The X-ray diffraction (XRD) is a powerful tool widely used in researches and the industry. While XRD is usually well known for qualitative and quantitative analyses of crystalline phases in materials, 
far more information can be obtained from a careful analysis of the diffraction
patterns or by using specific XRD settings: i.e. characterization of solid solutions, crystallite size and shape, crystal orientation, etc.
In XRD analyses, the diffraction of a wave of the characteristic length $\lambda$ is required to be of the same size as the interatomic separation of the crystal. However, XRD are in fact can also detect the inter-structure of a specific  atom which is called the small-angle techniques. 
The sketch map is shown in the Fig.~\ref{fig2:xrd}. Such small-angle technique, 
which is also known as the forward scattering in the scattering theory~\cite{PhysRevD.1.2349},
is a method of the investigation of the non-periodic systems. The physics principles of the scattering are the same for both wide-angle diffraction and the small-angle diffraction, in which the angle-dependent scattering amplitude is related to the electron-density distribution of the scattering by a Fourier
transformation. The main difference between wide- and small-angle diffraction in XRD
is that we have a periodic arrangement of identical scattering centers (particles)
in wide-angle diffraction, while the scattering center are limited in size,
non-oriented, and non-periodic, but the number of the particles is high and this 
can be assumed to be identical of the small-angle diffraction. Thus the small-angle diffraction can give a detailed investigation a single target. There are both coherent and incoherent scattering in the X-ray diffraction.
For the scattering on a single electron they are called Rayleigh scattering and Compton scattering, respectively.  
As shown in the Fig.~\ref{fig2:xrd},
the momentum transfer $q$ to the electron can be easily derived from the small-angle deviation from the incident X-ray, thus the parameter 
$\sin\theta/\lambda$ are always adopted for the description of the form factor in the momentum space with the relation
\begin{eqnarray}
 \frac{\sin\theta}{\lambda}=\frac{q}{4\pi}.
\end{eqnarray}
The summation of all the electrons gives the atomic 
scattering form factor
\begin{eqnarray}
 F(q)=\sum_j^{Z}f_j=\sum_j^{Z}\int\rho({\bm r}_j)\exp({\bm q}\cdot{\bm r}_j)
{\rm d}{\bm r}_j.
\end{eqnarray}
For the scattering by the atomic electrons, 
the phase of coherent scattering is by convention
related to that of the a free electron at the nucleus. There is a phase shift
of $\pi$ for scattering from a free electron.
And the coherent scattering in the XRD is~\cite{Prince:2016}
\begin{eqnarray}
 I_{\rm coh} = I_e\left(F(q)\right)^2.
\end{eqnarray}
In above equation, $I_e$ is the intensity of the
scattering per unit solid angle, it is calculated from the Thomson formula
\begin{eqnarray}
 I_e = I_0 r_e^2\left[\frac{1+2\cos^2\theta}{2}\right],
\end{eqnarray}
in which $I_0$ is the intensity of the unpolarized incident beam and $r_e$
is the classical radius of the electron ($2.818\times10^{-15}\rm m$). 
Note that $I_e$ and $I_0$ have different dimensions.
Another issue should be mentioned is that the momentum  of the 
incident X-ray $Q$ is much greater than the momentum transfer $q$.
This means that  the wavelength of incident X-ray are 
much smaller than the inverse momentum transfer, $\lambda_{X}\ll 1/q$. 
Thus above calculations are reliable in case of 
the wavelength of the X-ray larger than the classical radius of the electron,
no matter how large the atom is. This is the key point of the understanding
of the XRD which can be simply applied to the scattering of the dark matter.
The details are talked in the next section. 

It is usually considered as a common sense that the atomic form factor $F(q)$ 
could be described by a function equals the charge number $Z$ at zero momentum transfer and roughly decrease with the multiplicative inverse of the atom radius
quickly. However, things are contrary to the usual expectation.
For the Hydrogen atom, $F(q)$ can be got analytically by 
solving the  Schr\"odinger equation, it is effectively zero for 
$\sin\theta/\lambda>1.5{\mathring{\rm A}}^{-1}$. Those for heavier atoms are 
for relativistic wavefunctions which can be calculated by Relativistic 
Hartree-Fock (RHF) method of which the details can be found in Refs. \cite{Doyle:1968,Cromer:1968}
and all the results are summarized and listed in Ref.~\cite{Prince:2016}. 
Note that the RHF form factor is used for searching for Axion dark matter 
in inverse Primakoff scattering~\cite{Abe_2021} and in~\cite{Cadeddu_2019} trying to find observing atomic effects in coherent neutrino scattering. 
Usually, the form factor can be effectively described by ``Gaussian" fits
in the range $0{\mathring{\rm A}}^{-1}<(\sin \theta) / \lambda
<2.0 {\mathring{\rm A}}^{-1}$ 
\begin{eqnarray}
f(\sin \theta / \lambda)=\sum_{i=1}^{4} a_{i} \exp \left(-b_{i} \sin ^{2} \theta 
/ \lambda^{2}\right)+c\nonumber.
\end{eqnarray}
Take Xenon atom as an example, the fit parameters in above equation are
$a_{1}=20.2933,~b_{1}=3.9282;~a_{2}=19.0298,~b_{2}=0.344;
~a_{3}=8.9767,~b_{3}=26.4659;~
a_{4}=1.99,~b_{4}=64.2658;~c=3.7118$. But as the angle increases to 
$2.0{\mathring{\rm A}}^{-1}<(\sin \theta)/\lambda<6.0{\mathring{\rm A}}^{-1}$, 
Ref.~\cite{Fox:1989}  shows that the fitting formula above is highly
inaccurate, as the scattering angle increasing, 
and they produced a ``logarithmic polynomial" curve-fitting
routine based on the equation
\begin{eqnarray}
\ln \{f[(\sin \theta) / \lambda]\}=\sum_{i=0}^{3} c_{i} s^{i},
\end{eqnarray}
for these high angles. Also take Xenon atom as an example, the $c_{i}$ values are 
$c_{0}=4.24610,~c_{1}=-1.56330,~c_{2}=3.04200,~c_{3}=-2.34290$ 
give a close fit to the atomic scattering factor curves
over the range and the correlation coefficient is 0.999.

\begin{figure}[h]
\begin{center}
{\includegraphics[width=8cm]{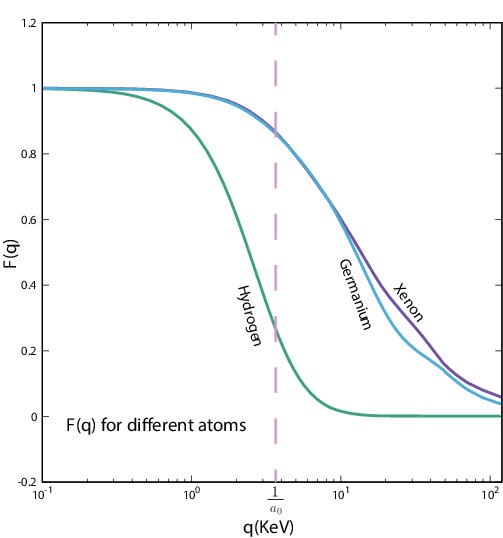}}
\caption{The atomic form factor of Hydrogen, Germanium and Xenon in which
the Hydrogen's is calculated analytically from Schr\"odinger Equation, and
the other two are calculated by Relativistic Hartree-Fock method. 
The corresponding momentum of inverse Bohr radius $1/a_0$ is shown in dash vertical line. All the form factors are normalized to 1 at zero momentum transfer 
$q=0$ for the comparison.}
\label{fig3:fq}
\end{center}
\end{figure}
In order to show the differences, we show the form factor $F(q)$ for
Hydrogen, Germanium and Xenon atoms in Fig.~\ref{fig3:fq},
in which the Hydrogen form factor is calculated analytically 
from Schr\"odinger Equation, and the other two are calculated by RHF.
The corresponding momentum of inverse Bohr radius $1/a_0$ is 
shown in dash vertical line in the panel. Note that all the form factors
are normalized to 1 at zero momentum transfer $q=0$ for the comparison.
The form factor of Hydrogen is a ``Dipole" form, and it decreases with the momentum transfer $q$ outside $1/a_0$ very quickly.
As for the Germanium (125pm)~\cite{doi:10.1063/1.1712084} and Xenon (108pm) 
of which the radii are at the order of Bohr radius, 
contrary to the common sense, the $F(q)$ does not decrease at 
the border of inverse radius. 
They can maintain a sufficiently large value to the much larger $q\gg 1/a_0$.
This is very important point of this work,  the large $F(q)$ at the
large momentum transfer can significantly enhance the rate of the recoils 
of the electron cloud. Before going to the detail, we show other
atomic form factor in the literature.

\subsection{Other Atomic Form factors}
In addition to RHF atomic form factor in XRD, there are several other form 
factors applied in different physical circumstances. 
The first one is the screened Coulomb form factor which is  usually
used for the detection of the Axion dark matter~\cite{Abe_2021}.
This form factor is defined for the atom of which the positive charge
of the nucleus in the atom center is screened by the electron 
cloud~\cite{PhysRevLett.103.141802, BUCHMULLER1990278, CEBRIAN1999397},
forming a Yukawa potential
\begin{eqnarray}
\phi(r)=\frac{Q}{4 \pi r} e^{-\frac{r}{r_{0}}},\label{yuka}
\end{eqnarray}
where $Q$ is the atomic charge, $r_0$ is the screening length.
In fact, Eq.~\eqref{yuka} is the electrostatic  potential of the atom.
Using Poisson's equation $\nabla^{2}\phi=-\rho$, 
we can get the charge density distributions outside the nucleus.
Note that the potential \eqref{yuka} is formed by the 
positive charge of the nucleus in the atom center ($\delta(r)$ function 
density) and the negative charge
of the electrons outside. Thus remove the delta function 
charge distribution and do the  Fourier transformation, 
we can obtain electron density distribution 
in the momentum space, namely the form factor
\begin{eqnarray}
F(q)=\frac{1}{1 + q^2 r_{0}^2}.\label{fqscreened}
\end{eqnarray}
For the  Xenon atom,  there are two results of the screening length used
in the literature: $r_{0}=0.49\mathring{\textrm{A}}$ is calculated in
Ref.~\cite{PhysRevLett.125.131806}, 
$r_{0}=2.45\mathring{\textrm{A}}$ is taken as the Wigner-Seitz radius 
in liquid Xenon~\cite{PhysRevLett.125.131805}. 

\begin{figure}[htbp]
\begin{center}
{\includegraphics[width=9cm]{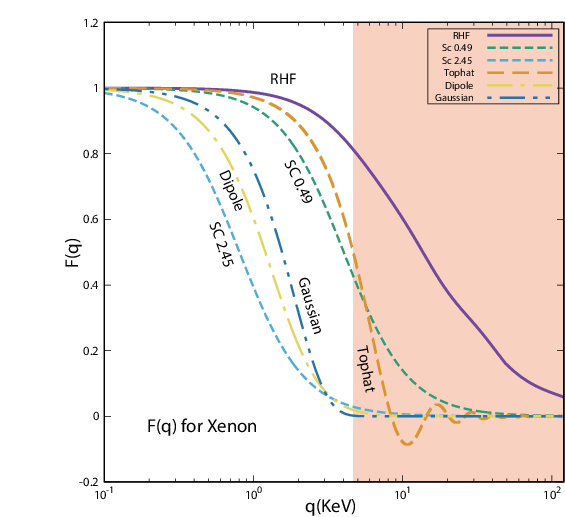}} 
\caption{Xenon's $F(q)$ gets from RHF method(purple line), screened Coulomb potential with two different radii(green/blue short dash) and the Tophat(orange dash), Dipole(yellow dot dash), Gaussian(blue double dots dash) shapes. The incarnadine shade is the interval of momentum transfer $q$ in the detection of light dark matter. 
The form factors are normalized to 1 at zero momentum transfer $q=0$ for the comparison.}
\label{fig4:fNq}
\end{center}
\end{figure}

As the inner structure of the electron cloud are rather complicate, there are three kinds of simple shapes, 
which are called ``Tophat", ``Dipole" and ``Gaussian", respectively. 
Those shapes are always used for the first exploration of the sized atom. The corresponding charge distribution $\rho (r)$,
root-mean-square radius $r_\text{Atom}$ and the form factor $F(q)$ are shown in following table
\\
\begin{center}
%\begin{table}
\begin{tabular}{cccc}\hline
Shape &  $\rho(r)$& $r_\text{Atom}$ & $F(q)$\\\hline
Tophat &$ \frac{3}{4\pi r_0^3} \theta(r_0-r)$ & $\sqrt{3/5}r_0$ 
& $\frac{3 (\sin (r_0 q)-r_0 q \cos (r_0 q))}{r_0^3 q^3}$  \\
Dipole &  $\frac{{\rm e}^{-r/r_0}}{8\pi r_0^3}  $  & $2\sqrt{3}r_0$
& $\frac{1}{\left(1+r_0^2 q^2\right)^{2}}$ \\
Gaussian  & $ \, \frac{1}{8 r_0^3 \pi^{3/2}} {\rm e}^{-r^2/(4r_0^2) }$ & $\sqrt{6} r_0$ & ${\rm e}^{-r_0^2q^2}$\\\hline
\end{tabular}
%\caption{Form factors for different density distributions. }
%\label{table:Fq}
%\end{table}
\end{center}
$r_0$ in the table is always chosen as the radius of the atom.
As talked above, the form factor of the ground state of the
Hydrogen atom is a  Dipole form.
The Helm form factor, which is always used to  describe the nucleon shape~\cite{PhysRevLett.128.101301},
is in fact  a Tophat form. These three kinds of form factors are also used in the 
puffy dark matter scenario  to realize the required velocity dependence in the small cosmological
scale~\cite{Chu:2020,Wenyu:2021}.

The RHF form factor used in XRD differs from the form factors shown in this subsection greatly. We show
the corresponding form factor $F(q)$ from  RHF result together
with the screened Coulomb (SC) potential,  Tophat shape etc. 
for Xenon atom in the Fig.~\ref{fig4:fNq}. 
Note that the radius $r_0$ for the Tophat, Dipole and Gaussian shapes is chosen as
$\thicksim2a_0$ which is got from Ref.~\cite{doi:10.1063/1.1712084}.
The two $r_0$ for the SC form factors shown in the figure are tagged
with ``SC0.49" and ``SC2.45", respectively.
The orange shade in Fig.~\ref{fig4:fNq} shows  the interval of momentum transfer 
$q$ for the detection of light dark matter which will be discussed in the next section. 
Another thing should be mentioned is that, the Tophat model describes a homogeneous
charge distribution within a sphere of radius $r_0$. We can see from the 
Fig.~\ref{fig4:fNq} and the analytical formula that $F(q)$ oscillate around zero from an
appropriate $q$. It means that the charge distribution in the
momentum space can be negative. This implies that the incident
particle can feel an opposite charge of electron cloud 
in case of a specific momentum transfer. This is a special phenomenon 
in the  effective scattering theory.

From Fig.~\ref{fig4:fNq}, we can see that the factors of 
the ordinary shapes quickly decrease to zero when $q$ is greater than the inverse Bohr radius $\sim 4\rm KeV$. 
Only the RHF form factor can maintain a large value to a much larger momentum. 
As discussed in  Ref.~\cite{Abe_2021}, the electrostatic potential cannot be expressed in a 
simple form Eq.~\eqref{yuka}, SC form factor Eq.~\eqref{fqscreened} is not the suitable
factor either. RHF form factor are more reliable for the realistic applications.
This implies that the  atoms like Xenon have non-trivial shapes for the electron outside the nuclei. 
$F(q)$ plays a very important role in the  scattering on the 
electrons in an atom.  The slowly decreasing behavior of the RHF form factor
can make the electrons recoils coherently, thus the 
cloud can receive a large momentum transfer $q$ from the incident 
particles. This is contrary to the common sense that the cloud only 
receive the momentum transfer $q$ at the order of the inverse radius of the atom.
Thus the coherent scatting and incoherent scattering are discussed in the following subsection.

\subsection{Coherent and incoherent scattering}
In order to figure out whether coherently scattering can happen for the electron cloud,
here we briefly review on the differences between  coherent scattering and 
incoherent scattering~\cite{PhysRevLett.128.101301,PhysRevD.88.116005}. 
When an incoming plane wave incident on a radially symmetric potential, 
the outgoing wave state can be written as
\begin{eqnarray}
S|\bm{p}\rangle=|\bm{p}\rangle+\int {\rm d}^{3} \bm{p}^{\prime} \delta\left(E_{\bm{p}}-E_{\bm{p}^{\prime}}\right) \mathcal{M}\left(\bm{p}, \bm{p}^{\prime}\right)\left|\bm{p}^{\prime}\right\rangle.
\end{eqnarray}
Here $\mathcal{M}\left(\bm{p}, \bm{p}^{\prime}\right)$ is the scattering matrix elements and $S$ is the usual S-matrix. 
If the target is composed of  $N$ identical potentials,  and 
each  potential is displaced away from the origin to some location
$\bm x_i \ne 0$, S-matrix for each potential  will be transformed by a unitary matrix
$U=\text{exp}(-{\rm i}\bm{x}_i\cdot\bm{p})$. Then the plane wave scatters to 
\begin{eqnarray}
|\bm{p}\rangle\to U S U^{\dagger}|\bm{p}\rangle
=|\bm{p}\rangle+\int {\rm d}^{3} \bm{p}^{\prime} \delta\left(E_{\bm{p}}-E_{\bm{p}^{\prime}}\right) {\rm e}^{-{\rm i} \bm{x}_i \cdot \bm{q}} \mathcal{M}\left(\bm{p}, \bm{p}^{\prime}\right)\left|\bm{p}^{\prime}\right\rangle,
\end{eqnarray}
where the momentum transfer is $\bm{q}=\bm{p}^{\prime}-\bm{p}$. To the lowest order in the perturbation theory,
the final state of the scattering is given by summing over all the amplitudes
\begin{eqnarray}
|\bm{p}\rangle\to |\bm{p}\rangle+\sum_{i=1}^N\int {\rm d}^{3} \bm{p}^{\prime} \delta\left(E_{\bm{p}}-E_{\bm{p}^{\prime}}\right) {\rm e}^{-{\rm i} \bm{x}_i \cdot \bm{q}} \mathcal{M}\left(\bm{p}, \bm{p}^{\prime}\right)\left|\bm{p}^{\prime}\right\rangle,
\end{eqnarray}
Therefore the probability of the momentum state $|\bm{p}^\prime \rangle$ in
the final state is proportional to
\begin{eqnarray}
P(\bm p^\prime) \propto \left|\langle\bm p^\prime|\psi\rangle\right|^2\propto\sum_{i j}\left\langle {\rm e}^{-{\rm i}\Delta\bm{x}_{i j} \cdot \bm{q}}\right\rangle\left|\mathcal{M}\left(\bm{p}, \bm{p}^{\prime}\right)\right|^{2},\label{prob}
\end{eqnarray}
where $\Delta\bm{x}_{i j}=\bm{x}_{i}-\bm{x}_{j}$. The expectation value is taken
over the N-body internal state of the target.  The circumstance is similar to the
light diffraction which occurs only when the slit is physically the approximate
size of, or even smaller than that light’s wavelength. 
For the N-body target, only when the momentum transfer is smaller than 
the inverse spacing between the potentials, i.e. 
$\Delta\bm{x}_{i j}\cdot\bm q<1$ for all $i,j$, that every phase in the sum 
is approximately the same, the scattering will be enhanced coherently. If the
spacing are too large, that only $N$ diagonal $i=j$  
elements in Eq.~\eqref{prob} can  survive. Then the incident particle
scattering on the each target separately, which is incoherent scattering. 

\begin{figure}[htbp]  
\includegraphics[width=10cm]{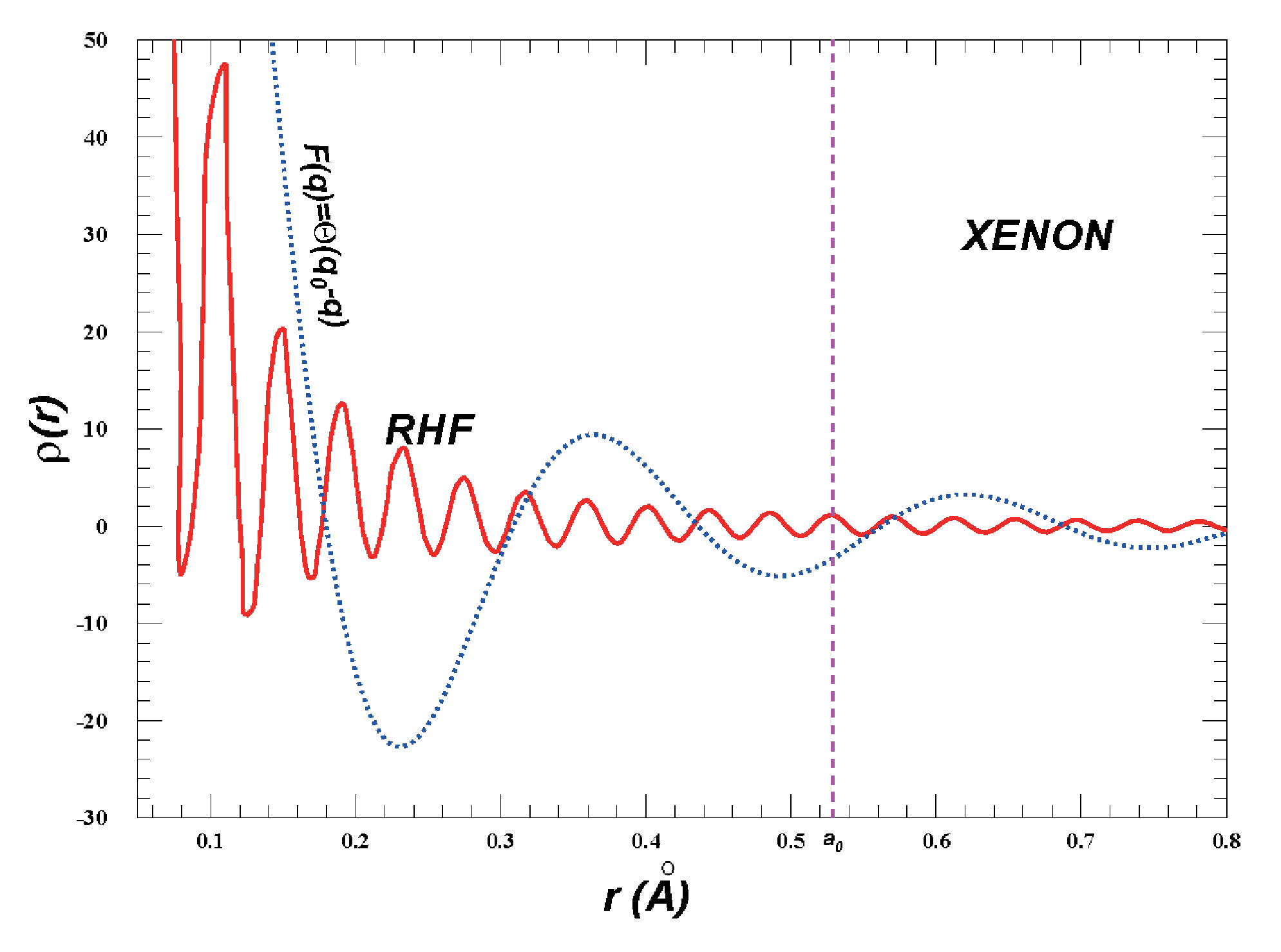}  
\caption{Charge distribution of $\rho(r)$ of Xenon, calculated by the Fourier
transformation of RHF $f(q)$ (red solid line) 
and a ``Tophat" model (blue dot line) in momentum space that
$F(q)=\theta(q_0-q)$ with $q_0=25\rm KeV$. The dash line shows the position of 
Bohr radius.}
\label{fig4:rr}
\end{figure}

From the argument above, we can see that the coherence depends on 
the space scale. As for the electrons in an atom, whether they can
behave as a compact ball at a specific momentum transfer $q$ depends
on the space scale where the  electron cloud locates. Naively thinking,
the scale would be the radius of the atom. However, as shown in 
last subsection, real electron clouds outside the nuclei have  
non-trivial shapes. The RHF form factor $F(q)$
does not vanish outside the inverse radius of the atom, but
decreases slowly with the momentum transfer $q$. In order to show the subtlety, 
we choose a simple form factor $F(q)$
\begin{eqnarray}
 F(q)=\theta(q_0-q),
\end{eqnarray}
which is a homogeneous charge distribution in the momentum space.
$q_0$ is the border of the distribution, and it is also the typical decreasing point of $F(q)$.
In fact this is  just  a ``Tophat" shape in the momentum space.
Obviously the $F(q)$ decreases to zero at $q=q_0$ in  this kind of  distribution. However,
the root-mean-square radius is not at the order of $1/q_0$, This can be 
derived from the shape of the charge distribution in coordinate space which is similar
\begin{eqnarray}
 \rho(r)=\frac{\sin (q_{0} r) - q_{0} r\cos (q_0 r)  }{ 2\pi^{2} r^3 }.\label{qrho}
\end{eqnarray}
The corresponding $\rho(r)$ is calculated by the Fourier transformation
of $F(q)$, shown in the Fig.~\ref{fig4:rr}, in which we chose $q_0=25\rm KeV$ for the demonstration.
Note that the unit charge of the target implies $F(q=0)=1$, thus the normalization factor
of $\rho(r)$ in Eq.~\eqref{qrho} is different from that in the table above.
The $\rho(r)$ of Xenon atom numerically calculated from RHF form factor $F(q)$ 
is shown in the red solid line of the figure.  
We can see that the negative charge distribution phenomenon discussed in above subsection come out again.
This is interesting  and  they can be considered as the effective charge distributions.
It means that the incident particle can feel the opposite charge effectively in the coordinate space. 
From the figure we can get the following implications. It is true that the 
charge density decreases quickly outside the typical inverse momentum $1/q_0$ in the Tophat model, 
while the distribution of the RHF is similar. However, 
this does not conflict with the measured radius of the atom.
We can see that the charge density can maintain the oscillating
behavior to the large radius. 
One can easily check that
the root-mean-square radii in both cases are oscillating divergent functions. 
The results imply that the charge is dominantly located in a very small region near the center of the atom,
The requirement of the coherent scattering $\Delta\bm{x}_{i j}\cdot\bm q<1$ can be satisfied.
Thus the electrons form a compact ball which can receive large momentum transfer $q$ and behave like a single particle.
Then the non-trivial shape can make the form factor much different from the ordinary consideration. 

From the Fig.~\ref{fig4:fNq}, we can see that Xenon RHF form factor $F(q)\sim 0.2$ at $q=50\rm KeV$. 
The physics picture for the the X-ray diffraction on this momentum transfer
is that the momentum transfer $q$ is much larger than inverse atom radius $1/r_0$. 
Since it is a small angle (forward) scattering $Q\gg q$, the wavelength of X-ray
is much smaller than the radius of the atom. This means that
a much smaller X-ray photon which can inject into the interior of the atom, 
can still transfer a much larger momentum to the whole atom and strongly 
interact with it. Whether such physics picture can happen in case of an incident dark matter? 
This will be discussed in the next section.

\section{Recoil of the electron cloud and its ionization}\label{sec3:th}
Compare to the XRD, the key difference in the scattering between the dark matter 
and the atom may be that the incident particle is massive and slowly moving.
However, our focus is the sub-GeV dark matter with a velocity as several hundred
kilometers per second ${\cal O} (10^{-3}c)$. Then the de Broglie wave length
for the dark matter lighter than 100 MeV can be much larger than the classical
radius of the electron. If the interaction between the dark matter and electron
is compact, the time scale of the momentum transfer will also be much
quicker than the light-crossing time of the electron cloud. 
The only problem is that the light-crossing time of the slowly 
moving dark matter is much longer than the X-ray, 
then non-trivial collision may happen.  
However, the whole picture is that whether the electron cloud can react
as a compact ball in the collision process. Fig.~\ref{fig4:fNq} shows that such compactness can only maintain
at a much smaller momentum transfer regime $q<1/r_0$ for the ordinary
Tophat, Dipole,  Gaussian or the SC shapes. But the RHF
results show that the compactness could be reliable to a much larger
$q$. If the compactness maintains, the longer collision time implies the
collision can be considered as to have higher order contributions from
the loop diagram, thus the amplitude must be at sub-leading order.
In all, the collision between the dark matter and the electron
electron cloud is analogous to the XRD scattering. The collision between dark matter
and the electron can transfer a much larger momentum to the whole electron 
cloud. The RHF form factor $F(q)$ could be used for the estimation
of the momentum transfer and the cross section.
In the following, we will give a detailed study on this recoil of 
electron cloud (REC) to see whether such estimation is suitable 
for the exploration of the dark matter.

In order to  compare the Electron ionization via  of both the Migdal effect and recoil of a single electron,
Ref.~\cite{Baxter:2019pnz} proposed the simple $U(1)$ extension model of the dark matter which is 
consist of  a dark matter candidate $\chi$ and dark photon $A'$. The $A'$ kinetically mixes with
the visible photon. In the work,  electron ionization from REC is calculated and compared  with the other
two processes. Thus the corresponding calculation schemes, such as the Lagrangian, kinetics and dynamics, are similar.
We just give some essential points and differences between  Ref.~\cite{Baxter:2019pnz}. Detailed calculation can be from  
Ref.~\cite{Baxter:2019pnz} and the references within.
First, the physical quantity for the comparison is denoted as 
\begin{eqnarray}
\bar\sigma_{e}=\frac{16\pi\epsilon^{2}\alpha\alpha_{D}\mu_{\chi e}^{2}}{(m_{A'}^{2}+\left|{q}_{0}\right|^{2})^{2}}.
\end{eqnarray}
It is the fiducial DM-free electron scattering cross section at fixed momentum transfer $q_0=\alpha m_{e}$, from which $\mu_{\chi e}$ is the reduced mass of the $\chi-e$ system. 
Next, we show the kinetics of the REC and compare it with the Migdal effect and recoil of a single electron.
In all these three processes, the incoming and outgoing states are the same: a dark matter particle plus a bound atom and a dark matter particle plus an ionized atom plus an unbound electron, respectively. 
The incoming dark matter is assumed to be a plane wave, which is both an energy eigenstate and a momentum eigenstate. The incoming atom (at rest in the lab frame) is an energy eigenstate and a momentum eigenstate too. 
The outgoing dark matter is also a plane wave, and the outgoing atom is in an excited state where the ionized electron belongs to the continuum spectrum of the atomic Hamiltonian. 
Consider the scattering process of a dark matter particle hitting electron clouds of $Z$ electrons outside the entire nucleus, for the dark matter particle with mass $m_{\chi}$, incoming velocity $\bm{\upsilon}$, 
and outgoing momentum $\bm{p}'_{\chi}$, momentum conservation requires
\begin{eqnarray}
\bm q\equiv m_{\chi}\bm{\upsilon}-\bm{p}'_{\chi}
\simeq Zm_{e}\bm{\upsilon}_{C}. \label{equation2}
\end{eqnarray}
Note that the recoil energy of the ionized electron 
is always negligible compared with the momentum transfer.
The energy conservation requires the ionized electron get  
\begin{eqnarray}
E_{e,f}-E_{e,i}\equiv  \Delta E_{e} = \frac{1}{2}m_{\chi}\upsilon^{2}-\frac{\left|m_{\chi}\bm{\upsilon}-\bm{q}\right|^{2}}{2m_{\chi}}-\frac{\bm{q}^{2}}{2Zm_{e}} = \bm{q}\cdot\bm{\upsilon}-\frac{\bm{q}^{2}}{2\mu_{\chi C}},
\end{eqnarray}
where $\mu_{\chi C}={Zm_{e}m_{\chi}}/{( Zm_{e}+m_{\chi})}$ is the DM-Cloud reduced mass. 

The kinematics of the Migdal effect, single electron 
and REC  scattering are similar, however, the dynamics differ from 
each other. They depend on the interacting objects, in other words, 
on which target the momentum are transferred to, the nucleus, electron or the electron cloud.
The interaction term in the perturbation Hamiltonian of the collision 
between the dark matter and electron cloud with $Ze$ is
\begin{eqnarray}
H_{\rm int,~C}=-\int\frac{\rm d^{3}\bm{q}}{(2\pi)^{3}}{\rm e}^{{\rm i}\bm{q}
\cdot(-\bm{x}_{\chi}+\bm{x}_{c})}\frac{\mathcal{M}_{\chi C}(q)}
{4m_{\chi}Zm_{e}},
\end{eqnarray}
where $\bm{x}_{c}$ is the charge distribution position of the electron cloud 
and ${\cal M}_{\chi C}$ is the Lorentz-invariant matrix element for dark matter
scattering off electron clouds through 4-momentum transfer $q\approx(0,\bm{q})$. 
Note that the phase of the charge distribution position 
${\rm e}^{{\rm i}\bm{q}\cdot(-\bm{x}_{\chi}+\bm{x}_{c})}$ is similar to the
phase in Eq.~\eqref{prob} discussed in above section. 
Comparing with the recoil of a single electron, 
what we only need to do is replacing the corresponding reduced mass 
and interaction, etc.
As the total charge of the electron cloud is $Z$, it is obvious that 
the $\mathcal{M}_{\chi C}$ can be reduced to a similar
interaction as the recoil of a single electron
\begin{eqnarray}
 \left|\mathcal{M}_{\chi C}\right|^2 =Z^2|F_C(q)|^2
 \left|\mathcal{M}_{\chi e}\right|^2,
\end{eqnarray}
in which $F_C(q)$ is the normalized form factor talked in above section.
Thus the same reduced cross section $\bar \sigma_e$ can be defined
to evaluate both recoils of the single electron and the cloud. 
In fact this is the usual way to do the comparison between 
detection strategies and the target, i.e. $\sigma_p^{\rm SI}$ 
spin-independent cross section for different target nuclei. 
Note that RHF electron cloud form factor for DM-electron-cloud scattering is
used here. The reason for this is that the kinetic mixing between the 
dark and visible $U(1)$ gauge field implies that 
the interactions between the massive dark photon $A'$ 
and standard model fermion are proportional to visible $U(1)$ charge.
Thus same form factor can be used as shown in Ref.~\cite{Abe_2021}. 

After the cloud are suddenly boosted to specific velocity $v_C$, it will
be dragged by the static nucleus and throw away an ionized electron. 
As talked in above section, it is in fact a similar Migdal effect
but happens in different Galilean frame. Thus the Rate of dragging back 
can be written in similar formula
\begin{eqnarray}
R_{C}\propto \left|\left\langle\Psi_{f}|\Psi_{\bm{\upsilon}_C}\right\rangle \right|^{2}
\sim\left|\left\langle\psi_{f}\left|
{\rm e}^{{\rm i}\bm{q_{e}}\cdot\bm{x}}\right|\psi_{i}
\right\rangle \right|^{2}
\end{eqnarray}
where $\Psi_{\bm{\upsilon}_C}$ and $\Psi_{f}$
are the initial and final states of the wave
function of an electron cloud, and they are the single-electron states wave
functions. However, being different with the 
Migdal effect, $\Psi_{\bm{\upsilon}_C}$ is the cloud with a velocity $\bm{\upsilon}_C$,
and the $\Psi_f$ is the final state of the almost static atom.
\begin{eqnarray}
\bm{q}_{e}=m_{e}\bm{\upsilon}_{C}=\frac{{\bm q}}{Z}.\label{qerec}
\end{eqnarray}
The reason of this estimation is that a compact ball for the electron cloud 
are adopted in the whole picture,  all the electrons
in the cloud will move at the same velocity. 
Note that, though a similar $q_e$ is using in both Migdal effect 
and REC, the $q_e$ in Migdal effect is 
$\bm{q}_{e}\simeq\frac{m_e}{m_N}\bm{q}$, which is much smaller than $q_e$ in Eq.~\eqref{qerec}, and this will make the phenomenology much different between the two processes. The detail will be discussed in the next section.

The calculation of the ionization rate in REC is also similar to Migdal effect and recoil of single electron. We need to integrate over the momentum transfer 
$\bm{q}$ and the dark matter velocity $\bm{v}$, 
then sum over the final electronic states $\Psi_{f}$ weighted by a delta function enforcing energy conservation. The specific normalization method of 
summation of final states can be found in Ref.~\cite{Baxter:2019pnz}.

As talked above, the reduced cross section $\bar\sigma_e$ is adopted of 
the calculation of the scattering rate. In fact, the cross section 
is proposed under the impulse approximation which is related to the free-particle
scattering matrix element as the electron recoil spectrum per unit detector mass.
The scattering of the dark matter with a single electron can treat the target
electron as single-particle states of an isolated atom, 
described by numerical RHF bound wave functions. 
The velocity-averaged differential ionization cross section
for electrons in the $(n,l)$ shell.
\begin{eqnarray}
\frac{{\rm d}\left\langle \sigma_{\rm ion}^{nl}\upsilon\right\rangle}{{\rm d}\ln E_{e}}
=\frac{\bar{\sigma}_{e}}{8\mu_{\chi e}^{2}}
\times\int q{\rm d}q \left|f_{\rm ion}^{nl}(k_{e},q)\right|^{2} \left|F_{DM}(q)\right|^{2}
\eta(\upsilon_{min}),
\label{eqn:123}
\end{eqnarray}
where $\eta(\upsilon_{min})=\left\langle
\frac{1}{\upsilon}\theta(\upsilon-\upsilon_{min})\right\rangle $ is 
the inverse mean speed for a given velocity distribution as a function of the
minimum velocity. $\upsilon_{min}$ is the minimum velocity required for
scattering. It can be solved by the delta function for energy conservation, going straight forward to get
\begin{eqnarray}
\upsilon_{min}=\frac{\left|E_{nl}\right|+E_{e}}{\left|\bm{q}\right|}+\frac{\left|\bm{q}\right|}{2\mu_{\chi C}}.
\end{eqnarray}
We assume a standard Maxwell-Boltzmann velocity distribution with
circular velocity $\upsilon_{0} = 220 \rm km/s$ and a hard cutoff 
$\upsilon_{\rm esc} = 544 \rm km/s$. The $q$-dependence of the matrix element
is encoded in the dark matter form-factor $F_{\rm DM}(q)$.
The differential ionization rate is written as
\begin{eqnarray}
\frac{{\rm d}R_{\rm ion}}{{\rm d}\ln E_{e}}=N_{T}\frac{\rho_{\chi}}{m_{\chi}}\sum_{nl}\frac{{\rm d}\left\langle \sigma_{\rm ion}^{nl}\upsilon\right\rangle }{{\rm d}\ln E_{e}},
\end{eqnarray}
where $N_{T}$ is the number of target atoms and $\rho_{\chi}=0.4\rm GeV/cm^{3}$ 
is the local dark matter density.

The key differences between the REC and the other two processes 
entirely hide in the ionization form factors, which are independent of 
all dark matter properties and depend only on the electronic 
and nuclear structure of the target. For coherent electron 
clouds scattering, the analogous ionization form factor is 
\begin{eqnarray}
\left|f_{\rm ion,~C}(E_{e}, q)\right|^{2}
=\frac{k_e^{3}}{4\pi^{3}}Z^{2}\left|F_{C}(q)\right|^{2}\times2\sum_{n,l,l',m'}\left|\left\langle \psi_{E_{e}}^{f}\left|{\rm e}^{{\rm i}\bm  {q_{e}}\cdot {\bm x}}
\right|\psi_{E_{nl}}^{i}\right\rangle \right|^{2}.
\end{eqnarray}
The other two ionization form factor can be found in Ref.~\cite{Baxter:2019pnz}.
First there is an additional $Z^2$ factor compared with the factor of the
recoil of a single electron. This is similar to the coherent enhancement in spin-independent scattering of the nucleus.
Another point is that $q_e$ in REC will be much larger that the $q_e$ in Migdal effect.

\begin{figure}[htb]
\begin{center}
{\includegraphics[width=8cm]{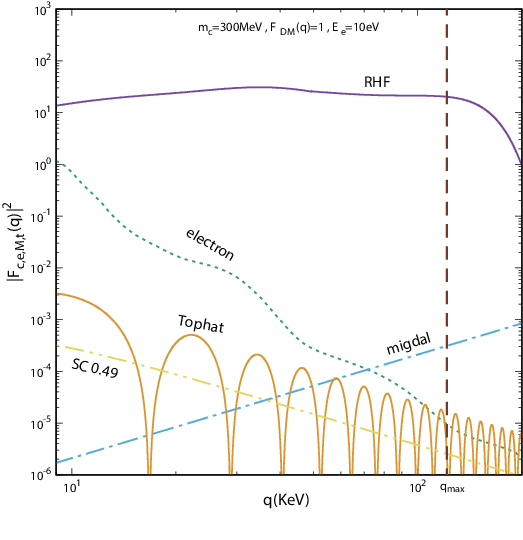}} 
\caption{Comparison with
$\left|f_{\rm ion,~e}\right|^2$ (green dot),  
$\left|f_{\rm ion,~M}\right|^2$ (sky blue dash dot) and 
$\left|f_{\rm ion,~C}\right|^2$ of the Xenon atom. For the ionization factor
of the electron cloud, the purple line shows the results of RHF form
factor, the orange line shows the result of the ``Tophat", and the yellow double dot dash line leads to 
the screened Coulomb potential with the screening length $r_{0}=0.49\mathring{\textrm{A}}$.
The maximum $q$ for the REC is shown in brown dash line.}\label{fig6:fq}
\end{center}
\end{figure}
The numerical calculation of the ionization factor for the Xenon atom
can be found in  Ref.~\cite{Baxter:2019pnz}. Here we 
chose $m_\chi = 300\rm MeV$, $E_e=10 \rm eV$ and $F_{\rm DM}=1$ and 
show the differences between the three processes in the Fig.~\ref{fig6:fq}.
In order to show the importance of the RHF electron cloud form factor,
the results of the Tophat shape and SC0.49 are also shown in the figure. 
The interval of $q$ is just the shaded region in the Fig.~\ref{fig4:fNq}. From Fig.~\ref{fig6:fq}, we can see that, the RHF form factor gives a much larger ionization form factor than the Migdal effect and the recoil of a single electron. However, the ionization factor of the Tophat shape and the screened Coulomb potential are much smaller than the recoil of the electron.
These two factors can be  at the same level with the Migdal effect. 
As we concern, these two models coarsely depict the electron coherent scattering, which are used only as references. The key point is that REC could dominate the ionization if we take the RHF form factor seriously. Another three points should be addressed: 
\begin{enumerate}
    \item As shown in the Fig.~\ref{fig6:fq}, $\left|f_{\rm ion,~e}\right|^2$
    decreases quickly with the $q$, thus the rate of recoil of a single electron 
    is dominated by small $q$ region.   
    \item $\left|f_{\rm ion,~M}\right|^2$ increases with the growing of $q$,
    thus the rate is dominated by large $q$ region. Nevertheless, as the 
    $m_\chi$ we considered is much lighter than the mass of the nucleus, the
    reduced mass of the collision mainly proportional  to $m_\chi$. Since
    $q_e$ in this effect are much smaller than $q$, the ionization factor
    does not feel any suppression factor in the region of the integration.
    \item The above two items are already addressed in Ref.~\cite{Baxter:2019pnz},
    the RHF $\left|f_{\rm ion,~C}\right|^2$ is larger than the other two 
    ionization factors. However, the upper limit of $q$ is 
    \begin{eqnarray}
    q_{\rm max}=2\mu_{\chi C}\upsilon_{\rm max}  \simeq 127.43\rm KeV,
    \end{eqnarray}
    for a dark matter much heavier than the mass of the Xenon cloud.
    Note that $\upsilon_{\rm max}$ is taken as usual $770\rm km/s$ 
    in above equation.
\end{enumerate}
With these ionization form factors, we can calculate the scattering rate
and get the limits on the scattering cross section which will be discussed
in the next section.

\section{ Numerical limits from the experimental detection}\label{sec4:nr}
Following the procedure in Ref.~\cite{Essig:2012yx}, 
we numerically calculate the limits on the spin-independent scattering
cross section from the experimental XENON100 data \cite{XENON100:2011cza,
XENON:2016jmt} (30 kg-years) and 
XENON1T data \cite{XENON:2019gfn} (1.5 tones-years). 
The recoil energy $E_e$ can induce
both of the observed electrons $n_{e}$ and scintillation photons $n_{\gamma}$.
The numbers of quanta $n^{(1)}$ that are produced by the step energy $W=13.8$~eV
which is  $n^{(1)} = {\rm Floor}(E_{\rm R}/W)$. The probability of the initial
electron recombining with an ion is assumed as $f_R=0$. 
The fraction of the primary quanta observed as the electrons 
is taken as $f_e = 0.83$.  The corresponding uncertainties are chosen
as $0<f_{R}<0.2$, $12.4<W<16$ eV, and $0.62<f_{e}<0.91$.
The photons are assumed from the deexcitation of the next-to-outer shells
$(5s,~4d,~4p,~4s)$, whose energies correspond to $(13.3,~63.2,~87.9,~201.4)$ eV. 
Due to the photoionization effect, these photons can respectively create an
additional quanta numbers $n^{(2)} = (n_{5s},~n_{4d},~n_{4p},~n_{4s}) = (0,~4,~6-10,~3-15)$ as shown in the Table~\ref{tabqe}
\cite{Essig:2017kqs}.  The electron event $n_{e}$ 
can be obtained by binomial distribution. The ionized electrons will convert
into the photoelectron (PE). The PE number is produced by an event of 
$n_e$ which obey ``Gaussian" distribution with mean $n_{e}\mu$ and width
$\sqrt{n_{e}\sigma}$,  where $\mu = 19.7 (11.4)$ and $\sigma = 6.2 (2.8)$ 
for XENON100 (XENON1T). Note that the data in the  range of [165, 275] of
the XENON1T data are adopted for our numerical evaluation.
\begin{table}[ht]
	\begin{tabular}{| l || l | l | l | l | l |}
		\hline 
		shell& 5$p^{6}$ & 5$s^{2}$ & 4$d^{10}$ & 4$p^{6}$ & 4 $s^{2}$\\
		\cline{1-6}
		Binding Energy[eV] & 12.6 & 25.7 & 75.6 & 163.5 & 213.8\\	
		\cline{1-6}
		Additional Quanta & 0 & 0 & 4 & 6-10 & 3-15\\
		\cline{1-6}
		\hline
	\end{tabular}
\caption{Additional quanta of the ionization of Xenon.}\label{tabqe}
\end{table} 

\begin{figure}[htpb]   
\begin{center}
{\includegraphics[width=16cm]  {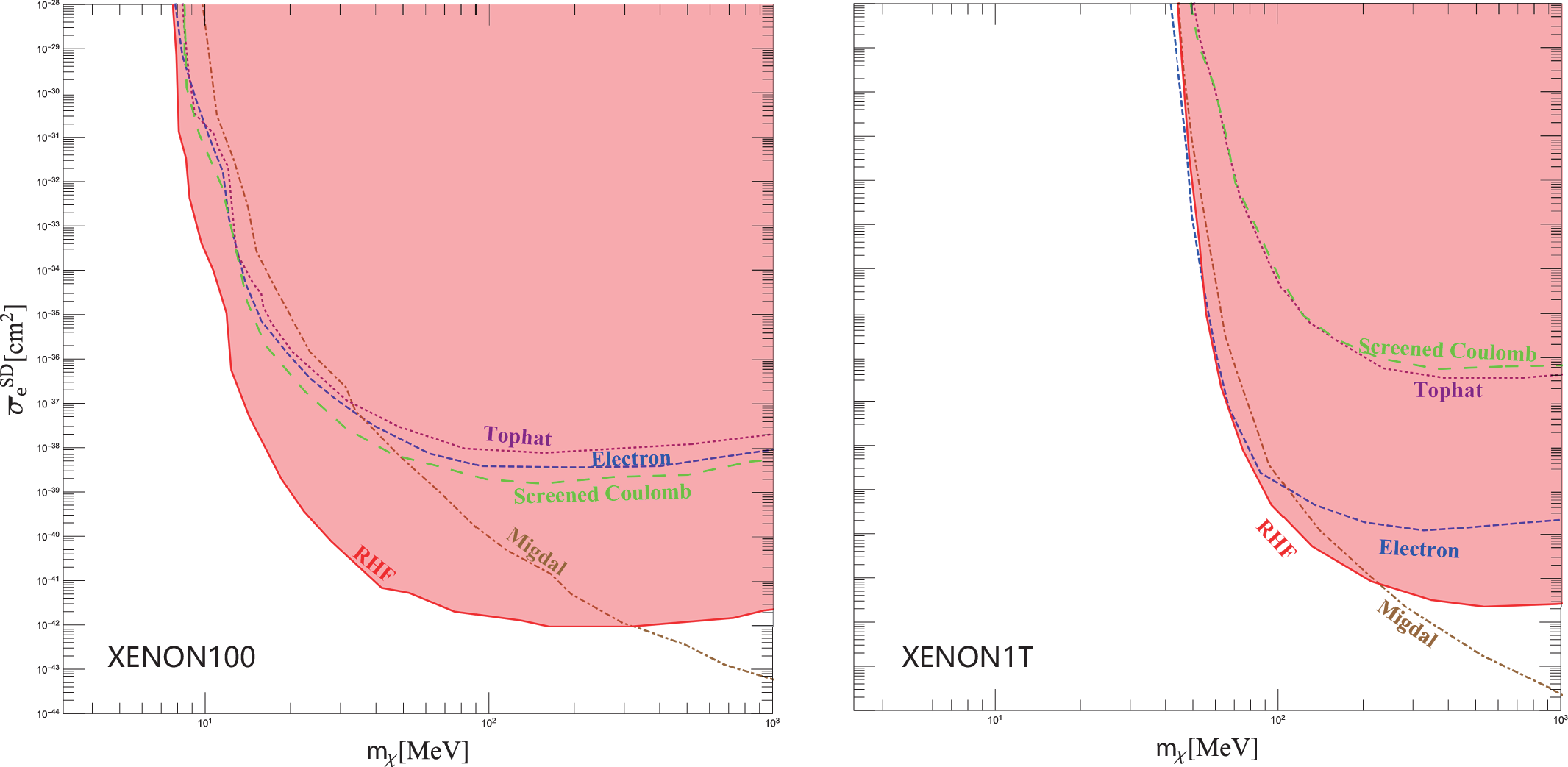}} 
\caption{The limits of the  cross sections  and dark matter mass
from Xenon100 (left panel) and Xenon1T(right panel).
The blue dash  (dot dash) lines show limits on the recoil of single electron and the Migdal effect. The dot line shows the result of Tophat model of the electron cloud, while the long green dash line gives the exclusion area from the screened Coulomb potential model with the screening length $r_{0}=0.49\mathring{\textrm{A}}$. The shaded regions show results of RHF form factor.}\label{fig6:lim}
\end{center}
\end{figure}

\begin{figure}[htpb]   
\begin{center}
{\includegraphics[width=16cm]  {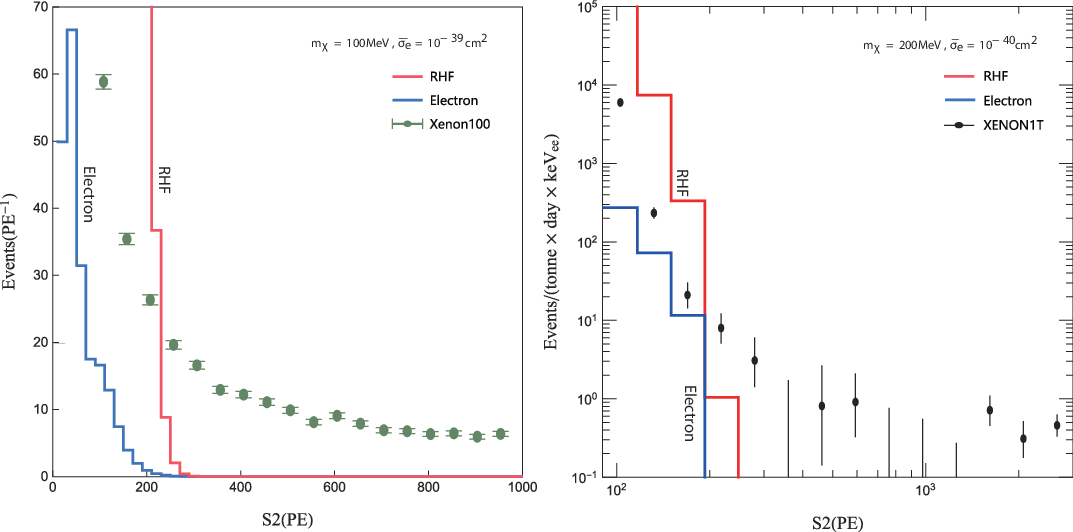}} 
\caption{Observed events versus photonelectrons (PE)
from Xenon100 (left panel) and Xenon1T (right panel) 
data, for Xenon100 $\bar \sigma_e = 1\times 10^{-39} \rm cm^{2}$ and $m_\chi = 100\rm MeV$ and while for Xenon1T, $\bar \sigma_e = 1\times 10^{-40} \rm cm^{2}$ and $m_\chi = 100\rm MeV$.}\label{fig8:pe}
\end{center}
\end{figure}

Fig.~\ref{fig6:lim} shows the numerical results of the limits from 
Xenon100 (left panel) and Xenon1T (right panel) data. The results
of limits on the recoil of a single electron and Migdal effect are
also shown for the comparison. We can see that, in the range of
$m_\chi \gsim 100\rm MeV$, the limits of the RHF form factor are 
much stringent than the recoil of a single electron,
almost 4 orders stronger. This can be easily understood
that the coherent enhancement of $Z^2 (54^2)$ factor and a sufficiently large
RHF $F(q)$ in the interval of the rate integral Eq.~\eqref{eqn:123}.
The limits on the Tophat model and screened Coulomb potential model 
are at the same order of the results from the recoil of single electron under Xenon100 experiment data. However, they are negligible to the recoil of single electron under the constraints of Xenon1T experiment data.
The limits on RHF form factor in both cases are much stringent than
the limits of recoil of single electron, Tophat and screened Coulomb potential, due to the RHF $F(q)$.
Note that, we can't distinguish all these three processes (REC, single electron and Migdal effect) in experiments now, because the number of PE generated is observed in the experimental results, and these three processes cause the electron to ionize out and generate corresponding numbers of photoelectrons as S2 signals. It is unable to tell which physical process produces these photoelectrons. Since recoil of a single electron and REC originate from the same dark $U(1)$ interaction, thus we just added together the both recoils, so that the summation gives a more comprehensive limits on the cross sections.
The constraints of the Xenon1T data released below the 100MeV, however,
the Xenon100 data can maintain the constraints to about 10MeV.
The reason for this can be shown in Fig.~\ref{fig8:pe}  which 
shows the observed events versus photonelectrons (PE) from 
the Xenon100 and Xenon1T experimental data.
The left panel shows the $\bar \sigma_e = 1\times 10^{-39} \rm cm^{2}$ and $m_\chi = 100\rm MeV$
result for Xenon100, and the right penal shows the   
$\bar \sigma_e = 1\times 10^{-40} \rm cm^{2}$ and $m_\chi = 200\rm MeV$results for 
Xenon1T. One can check that, 
the lighter dark matter ($<100 \rm MeV$) in the recoil of a single electron and
REC mainly cause different distribution in small PE number regions.
In the numerical calculation we choose [80,~190] PE bins 
of the Xenon100 for the constraints,
while we only choose 165 PE as the bin of the Xenon1T for the constraints.
Thus the Xenon100 give a better performance when the dark matter
mass lighter than 100 MeV.

As for the comparison with Migdal effect, the limits on the RHF form factor
are more stringent than the Migdal effect below about 400MeV (200MeV)  
from Xenon100  (Xenon1T) data. The reason has been addressed above that
the upper limit of the rate integral Eq.~\eqref{eqn:123} $q_{\rm max}$
in the Migdal effect can extend to a much larger value
when dark matter mass is much lighter than
the target nucleus. The ionization form factor can grow up to a sufficiently large value. Thus, the Migdal effect give a better performance
in heavy dark matter region.

\section{Conclusion}\label{sec5:con}
The searching for sub-GeV dark matter is a hot topic 
at the cutting edge of physics research. However, as the sensitivity
lost for the light dark matter detection, 
the traditional strategies for detecting WIMP-type dark
matter are no longer feasible. A novel detection are proposed
that the electron cloud is boosted by the dark matter 
and throws away an electron when it is dragged back by the heavy nucleus,
namely the coherent scattering of the electron cloud of the atom.
The processes can be considered as an similar Migdal effect but
in different Galilean fame. Though the form factor $F(q)$ decrease quickly with the multiplicative inverse of the radius of the atom, as shown in the simple Tophat, Dipole, Gaussian shape and the SC model, this effect is still not negligible compared with the recoil of a single electron and the Migdal effect. What's more, the survey in the X-ray diffraction show that the atomic form factor are much complicate than the naive consideration. The results
of the relativistic Hartree-Fock method gives non-trivial shapes of
the atom. We show that the corresponding form factors do not
conflict with measured radii in the coordinate space.
We also show that the collision between the sub-GeV dark matter and the 
electron cloud is similar to the X-ray diffraction. At least, 
the RHF form factor should be tried to be applied in the estimation of the scattering rate
of the dark matter detection. 
Having equipped with the RHF form factor and impulse approximation, we proceed to
show the detailed calculation of recoil of the electron cloud.
the kinetics, the fiducial cross section
and the corresponding calculation of detection rate are given analytically.
After that, we show the constraints on the cross section from the current
experimental measurements. The Xenon100 and Xenon1T data are adopted for the 
constraints. The comparison between the other two processes are also discussed.
We found that the limits of the RHF form factor are 
much stringent than the recoil of a single electron,
almost 4 orders stronger. The constraints of the Xenon1T data  
released below the 100MeV, however,
the Xenon100 data can maintain the constraints to about 10MeV.
The limits on the RHF form factor
are more stringent than the Migdal effect below about 400MeV (130MeV)  
from Xenon100  (Xenon1T) data, due to the 
extended  interval of the rate integration.
Another thing should be emphasized is that,
the recoil of a single electron 
and recoil of the electron cloud originate from the same fundamental interaction,
thus in principle both recoils should be added together and the summation 
gives a more comprehensive limits on the cross section. 

The process we proposed and the 
corresponding results seem promising and novel. However, 
the key point of our work is that whether the RHF form factor can be used
in the detection of the dark matter, which  we only have  neutral 
attitude on it.

\section{Acknowledgements}
We thank Nick Houston, Lei Wu and Bin Zhu for their helpful discussions. 
This work was supported by the Natural Science Foundation of China under 
grant number 11775012.

\bibliography{refs}
\end{document}